\documentclass[aps,twocolumn,showpacs,showkeys]{revtex4}

\usepackage{amsmath} \usepackage{graphicx} \usepackage{color}
\usepackage{amsfonts} \usepackage{hyperref}
\usepackage{palatino}\usepackage{gensymb}

\newcommand{\m}[1]{{\bf #1}} 
\newcommand{\gm}[1]{\mbox{\boldmath $#1$}}

\newcommand{\gerald}[1]{\textcolor{black}{#1}}
\newcommand{\Gerald}[1]{\textcolor{black}{#1}}

\usepackage[normalem]{ulem}

\begin{document}
\bibliographystyle{apsrev}
\begin{sloppy}
  \title{Protein secondary structure analysis with a coarse-grained
  model}

  \author{Gerald R. Kneller$^{1,2,3}$}
  \email{gerald.kneller@cnrs-orleans.fr} \author{Konrad Hinsen
    $^{1,2}$}
 
  \affiliation{$^{1}$Centre de Biophys. Mol\'eculaire, CNRS; Rue
    Charles Sadron, 45071 Orl\'eans,France}
  \affiliation{$^{2}$Synchrotron Soleil; L'Orme de Merisiers, 91192
    Gif-sur-Yvette, France} \affiliation{$^{3}$Universit\'e
    d'Orl\'eans; Chateau de la Source-Av. du Parc Floral, 45067
    Orl\'eans, France}
        
\begin{abstract}
  The paper presents a geometrical model for protein secondary
  structure analysis which uses only the positions of
  the $C_{\alpha}$-atoms. We construct a space curve connecting these
  positions by piecewise polynomial interpolation and describe the
  folding of the protein backbone by a succession of screw motions
  linking the Frenet frames at consecutive
  $C_{\alpha}$-positions. Using the ASTRAL subset of the SCOPe data
  base of protein structures, we derive thresholds for the screw
  parameters of secondary structure elements and demonstrate that the
  latter can be reliably assigned on the basis of a
  $C_{\alpha}$-model.  For this purpose we perform a comparative study
  with the widely used DSSP (Define Secondary Structure of Proteins)
  algorithm.
\end{abstract}

\pacs{87.15.-v, 87.15.B-, 87.15.bd}
 
\keywords{Protein secondary structure, coarse-grained protein model}

\vspace*{3 mm}

\maketitle

\section{Introduction}

Protein secondary structure elements (PSSE) are the basic building
blocks of proteins and their form and arrangement is of fundamental
importance for protein folding and function. They have been first
predicted by Pauling and Corey on the basis of hydrogen
bonding~\cite{Pauling:1951th,Pauling:1951ud} and were later confirmed
by X-ray diffraction experiments. The localization of PSSEs in protein
structure databases is one of the most basic tasks in bioinformatics
and various methods have been developed for this purpose. We mention
here DSSP (Define Secondary Structure of Proteins)\cite{Kabsch:1983wk}
and STRIDE (STRuctural IDEntification)~\cite{Frishman:1995tq}, which
assign PSSEs on the basis of geometrical, energetic and statistical
criteria and which are the most widely used approaches. The result are
contiguous domains along the amino acid sequence of the protein, which
are labeled as ``$\alpha$-helix'', ``$\beta$-strand'', etc. There is
no precise and universally accepted definition for PSSEs, and
therefore each method produces slightly different results. The
geometrical variability of these PSSEs, which depends on the global
protein fold, is not explicitly considered by these approaches. The
\gerald{more recently} published ScrewFit
method~\Gerald{\cite{Kneller:2006p13,Calligari:2012eja}} allows for
both assignment and geometrical description of PSSEs. It describes the
geometry of the protein backbone by a succession of screw motions
linking successive \mbox{$C-O-N$}~groups in the peptide bonds, from
which PSSEs can be assigned on the basis of statistically established
thresholds for the local helix parameters. The latter have been
derived by screening the ASTRAL database~\cite{Chandonia:2004eh},
which provides representative protein structure sets containing
essentially one secondary structure motif. The ScrewFit description is
intuitive and bears some ressemblances with the P-Curve approach
proposed by Sklenar, Etchebest and Lavery~\cite{Sklenar:1989vv}, in
the sense that both methods lead to a sequence of local helix axes,
the ensemble of which defines an overall axis of the protein under
consideration.  \gerald{ScrewFit} uses, however, a \gerald{minimal}
set of parameters and was originally developed to pinpoint changes in
protein structure due to external stress.
 
The experimental basis for the automated assignment of PSSEs in
proteins is X-ray crystallography, which yields information about the
positions of the heavy atoms in a protein. Although the number of
resolved protein structures increased almost exponentially during the
last two decades, the fraction of proteins for which the atomic
structure is known is still very small. Low resolution techniques,
like electron microscopy, are an additional source of information
\Gerald{\cite{Grimes:1999it,Marabini:2013bd}} and in this context the
description of PSSEs must be correspondingly simplified, in order to
be useful in structure refinement. A natural and commonly used
coarse-grained description of proteins is the $C_{\alpha}$-model,
where each residue is represented by its respective $C_{\alpha}$-atom
on the protein backbone~\cite{Tozzini:2005em}. To our knowledge,
Levitt \textit{et al.} were the first to publish a method of secondary
structure assignment on the basis of the
$C_{\alpha}$-positions\cite{Levitt:1977uu}, and different approaches
for that purpose have been published since
then~\cite{Dupuis:2004hp,Labesse:2005vt,Park:2011ji}. Like DSSP and
STRIDE, these methods aim at assigning PSSEs on a true/false basis and
the underlying models for this decision are not exploited or not
exploitable for a more detailed description of protein folds. The
motivation of this paper was to develop an extension of the ScrewFit
method which works only with the $C_{\alpha}$-positions, maintaining
the capability to describe the global fold of a protein by a
minimalistic model and to assign PSSEs. The method is described in
Section~\ref{sec:Method} and two applications are presented and
discussed in Section~\ref{sec:Applications}. A short r\'esum\'e with
an outlook concludes the paper.
 
\nopagebreak
 
\section{A coarse-grained model for the fold of a protein}
\label{sec:Method}

\subsection{$C_{\alpha}$ space curve and Frenet frames}
\label{sec:FrenetFrames}

We consider the ensemble of the $C_{\alpha}$-positions,
$\{\m{R}_{1},\ldots,\m{R}_{N}\}$, as a discrete representation of a
space curve, $\m{r}(\lambda)=\sum_{k=1}^{3}r_{k}(\lambda)\m{e}^{(k)}$,
where $\lambda\in[\lambda_{a},\lambda_{b}]$ and $\m{e}^{(k)}$
($k=x,y,z$) are the basis vectors of a space-fixed Euclidean
coordinate system. Imposing that
\begin{equation}
  \m{r}(\lambda_{j})=\m{R}_{j},\quad j=1\ldots N,
\end{equation}
at equidistantly sampled values of $\lambda$,
\begin{equation}
  \lambda_{j} = \lambda_{a}+ (j-1)\Delta\lambda,
  \quad\Delta\lambda =(\lambda_{b}-\lambda_{a})/N,
\end{equation}
we define a continuous space curve by a piecewise polynomial
interpolation of the $C_{\alpha}$-\gerald{positions}. The values for
$\lambda_{a}$ and $\lambda_{b}$ are arbitrary and one may in
particular choose $\lambda_{a}=0$ and $\lambda_{b}=N$, such that
$\Delta\lambda=1$.  \gerald{At each $C_{\alpha}$-position, we
  construct the local Frenet basis from the interpolated space curve},
\begin{align}
  \label{eq:FrenetContinuousT}
  \m{t}(\lambda)&=\frac{\dot{\m{r}}(\lambda)}{|\dot{\m{r}}(\lambda)|},
  \\\noalign{\medskip}
  \label{eq:FrenetContinuousN}
  \m{n}(\lambda)&=\frac{\dot{\m{t}}(\lambda)}{|\dot{\m{t}}(\lambda)|},
  \\\noalign{\medskip}
  \label{eq:FrenetContinuousB}
  \m{b}(\lambda)&=\m{t}(\lambda)\wedge\m{n}(\lambda),
\end{align}
where $\{\m{t},\m{n},\m{b}\}$ are, respectively, the tangent vector,
the normal vector, and the bi-normal vector to the curve. The dot
denotes a derivative with respect to $\lambda$. Interpolating the
space curve around each $C_{\alpha}$-position with a second order
polynomial involving the respective left and right neighbors, we
obtain
\begin{align}
  \dot{\m{r}}(\lambda_{j})&=\frac{\m{R}_{j+1}-\m{R}_{j-1}}{2\Delta
    \lambda}, \\\noalign{\medskip}
  \ddot{\m{r}}(\lambda_{j})&=\frac{\m{R}_{j+1}-2\m{R}_{j}+
    \m{R}_{j-1}}{\Delta \lambda^{2}},
\end{align}
for $j=2,\ldots,N-1$. At the end points of the chain one can only use
forward and backward differences, respectively, and a second-order
interpolation of the $C_{\alpha}$-space \gerald{would lead} to
identical $\{\m{t},\m{n}\}$-planes at the first and last two
$C_{\alpha}$-positions, which is not compatible with a helicoidal
curve. In this case we resort to third-order interpolation, such that
\begin{align}
  \label{eq:third-order-1}
  \dot{\m{r}}(\lambda_{1})&= \frac{-11 \m{R}_1+18 \m{R}_2-9 \m{R}_3+2
    \m{R}_4}{6 \Delta \lambda }, \\\noalign{\medskip}
  \label{eq:third-order-2}
  \ddot{\m{r}}(\lambda_{1})&= \frac{2 \m{R}_1-5 \m{R}_2+4
    \m{R}_3-\m{R}_4}{\Delta \lambda^{2} },\\\noalign{\medskip}
  \label{eq:third-order-3}
  \dot{\m{r}}(\lambda_{N}) &= \frac{-2 \m{R}_{N-3}+9 \m{R}_{N-2}-18
    \m{R}_{N-1}+11 \m{R}_N}{6 \Delta \lambda}, \\\noalign{\medskip}
  \label{eq:third-order-4}
  \ddot{\m{r}}(\lambda_{N})&= \frac{-\m{R}_{N-3}+4 \m{R}_{N-2}-5
    \m{R}_{N-1}+2 \m{R}_N}{\Delta \lambda ^2}.
\end{align}
We note here that the Frenet frames constructed at the
$C_{\alpha}$-positions 2--$N$ are identical with the so-called
``discrete Frenet Frames'' introduced in Ref.~\cite{Hu:2011bx}.

\subsection{Relating Frenet frames by screw motions}
\label{sec:ScrewFrame}
Having constructed the Frenet frames, the next step consists in
constructing the screw motions which link consecutive frames along the
protein main chain. For this purpose, the basis vectors
$\{\m{t}(\lambda_{j}),\m{n}(\lambda_{j}),\m{b}(\lambda_{j})\} \equiv
\{\m{t}_{j},\m{n}_{j},\m{b}_{j}\}$ must be referred to their
respective anchor points, $\m{R}_{j}$. Defining
\begin{equation}
  \gm{\epsilon}_{j}^{(1)}=\m{t}_{j},\quad \gm{\epsilon}_{j}^{(2)}=\m{n}_{j},
  \quad\gm{\epsilon}_{j}^{(3)}=\m{b}_{j},
\end{equation}
the ``tips'' of the Frenet basis vectors are located at
\begin{equation}
  \m{x}_{j}^{(k)}=\m{R}_{j}+\gm{\epsilon}_{j}^{(k)}\quad(k=1,2,3),
\end{equation}
and the mathematical problem consists in finding the screw parameters
for the mappings $ \{\m{x}_{j}^{(k)}\}\to\{\m{x}_{j+1}^{(k)}\}$ for
$j=1,\ldots,N-1$.

\subsubsection{Screw motions}
In general, a rigid body displacement $\m{x}\to \m{y}$ can be
expressed in the form
\begin{equation}
  \label{eq:RBdisplacement}
  \m{y} = \m{x}^{(c)}+\m{D}\cdot(\m{x}-\m{x}^{(c)}) + \m{t},
\end{equation}
where $\m{x}^{(c)}$ is the center of rotation, $\m{D}$ is a rotation
matrix, and $\m{t}$ a translation vector. By construction,
\begin{equation}
  \m{t}=\m{y}^{(c)}-\m{x}^{(c)}.
\end{equation}
The elements of the rotation matrix can be expressed in terms of three
independent real parameters. One possible choice is to use the
rotation angle, $\phi$, and the unit vector, $\m{n}$, pointing into
the direction of the rotation axis. For this parametrization, $\m{D}$
has the form\cite{Altmann:1986}
\begin{equation}
  \label{eq:DnPhi}
  \m{D}(\m{n},\phi)  = \cos\phi\,\m{1}
  +(1-\cos\phi)\,\m{P}+\sin\phi\,\m{N},
\end{equation}
where $\m{P}=(n_{i}n_{j})$ ($i,j=1,2,3$) is the projector on $\m{n}$
and $\m{N}$ is a skew-symmetric $3\times 3$ matrix which is defined by
the relation $\m{N}\cdot\m{v}=\m{n}\wedge\m{v}$ for an arbitrary
vector $\m{v}$. The elements of $\m{N}$ are
$N_{ij}=-\sum_{k}\epsilon_{ijk}n_{k}$, where $\epsilon_{ijk}$
($i,j,k=1,2,3$) are the components of the totally antisymmetric
Levi-Civita tensor. We recall that $\epsilon_{ijk}=\pm 1$ \gerald{for,
  respectively, an even and odd permutation of 123}, and
$\epsilon_{ijk}=0$ zero otherwise.  The parameters of the rigid-body
displacement~(\ref{eq:RBdisplacement}) depend on the choice of the
rotation center, $\m{x}^{(c)}$, and there is a special choice,
$\m{x}^{(c)}=\m{s}$, for which the translation vector $\m{t}$ points
into the direction of the rotation axis $\m{n}$, such that
$\m{t}\cdot\m{n}>0$. This is known as Chasles'
theorem~\cite{Chasles:1830} and the corresponding rigid body
displacement describes a screw motion,
\begin{equation}
  \label{eq:RBscrew}
  \m{y} = \m{s}+\m{D}(\m{n},\phi)\cdot(\m{x}-\m{s}) + \alpha\m{n}.
\end{equation}
Using that $\m{D}(\m{n},\phi)\cdot\m{n}=\m{n}$, one shows easily that
$\alpha$ is the projection \gerald{of the translation vector on the
  rotation axis},
\begin{equation}
  \alpha=\m{t}\cdot\m{n}.
\end{equation}
The position $\m{s}$ is not uniquely defined, but stands for all
points on the screw axis. Defining $\m{s}^{(c)}$ to be the point for
which the distance $|\m{s}-\m{x}^{(c)}|$ is a minimum, the screw axis
is defined through
\begin{equation}
  \m{s}=\m{s}^{(c)}+\mu\m{n},\quad -\infty<\mu<+\infty,
\end{equation}
where
\begin{equation}
  \label{eq:Sperp}
  \m{s}^{(c)} = \m{x}^{(c)}+\frac{1}{2}
  \bigl(\m{t}^{\perp}+\cos(\phi/2)\m{n}\wedge\m{t}\bigr),
\end{equation}
and $\m{t}^{\perp}=\m{t}-(\m{n}\cdot\m{t})\m{n}$ is the component of
$\m{t}$ which is perpendicular to the rotation axis.  We note that
$(\m{s}^{(c)}-\m{x}^{(c)})\cdot\m{n}=0$. The radius of the screw
motion is defined through $\rho=|\m{x}^{(c)}-\m{s}^{(c)}|$ and it
follows from~(\ref{eq:Sperp}) that
\begin{equation}
  \label{eq:rho}
  \rho = \frac{|\m{t^{\perp}}|}{2}\sqrt{1+\cot(\phi/2)^{2}}.
\end{equation}

\subsubsection{Determining the screw parameters}
Assuming that the Frenet frames at the $C_{\alpha}$-positions have
been constructed, the fold of a protein is defined by the sequence of
screw motions $\m{x}_{j}^{(k)}\to \m{x}_{j+1}^{(k)}$, where
\begin{equation}
  \m{x}_{j+1}^{(k)}=\m{s}^{(c)}_{j}+\m{D}(\m{n}_{j},\phi_{j})
  \cdot(\m{x}_{j}^{(k)}-\m{s}^{(c)}_{j}) + \alpha_{j}\m{n}_{j},
\end{equation}
for $j=1,\ldots,n-1$ and $k=1,2,3$.  The corresponding parameters are
computed as follows:
\begin{enumerate}
\item\label{eq:Translations} Determine the translation vectors
  \begin{equation}
    \m{t}_{j}=\m{R}_{j+1}-\m{R}_{j}.
  \end{equation}

\item Perform a rotational least squares fit\cite{Kneller:1991p71}
  $\{\gm{\epsilon}_{j}^{(k)}\}\to\{\gm{\epsilon}_{j+1}^{(k)}\}$ by
  minimizing the target function
  \begin{equation}
    \label{eq:LeastSquaresQuat}
    m(Q_{j})=
    \sum_{k=1}^{3}\left|\gm{\epsilon}_{j+1}^{(k)}-\m{D}(Q_{j})
      \cdot\gm{\epsilon}_{j}^{(k)}\right|^{2}
  \end{equation}
  with respect to four quaternion parameters,
  $Q=\{q_{0},q_{1},q_{2},q_{3}\}$, which parametrize the rotation
  matrix according to
  \begin{equation}
    \label{eq:Dq}
    \m{D}(Q) = \footnotesize \left(
      \begin{array}{ccc}
        q_0^2+q_1^2-q_2^2-q_3^2 & 2 \left(q_1 q_2-q_0 q_3\right) & 2
        \left(q_0 q_2+q_1 q_3\right) \\
        2 \left(q_1 q_2+q_0 q_3\right) & q_0^2-q_1^2+q_2^2-q_3^2 & -2
        \left(q_0 q_1-q_2 q_3\right) \\
        -2 \left(q_0 q_2-q_1 q_3\right) & 2 \left(q_0 q_1+q_2 q_3\right) &
        q_0^2-q_1^2-q_2^2+q_3^2 \\
      \end{array}
    \right).
  \end{equation}
  The quaternion parameters are normalized such that
  $q_{0}^{2}+q_{1}^{2}+q_{2}^{2}+q_{3}^{2}=1$, which leaves three free
  parameters describing the rotation.  We note here only that the
  minimization of~(\ref{eq:LeastSquaresQuat}) leads to an eigenvector
  problem for the optimal quaternion, which can be efficiently solved
  by standard linear algebra routines, and that the corresponding
  eigenvalue is the squared superposition
  error~\cite{Kneller:1991p71}.  The latter is zero for superposition
  of Frenet frames, since two orthonormal and equally oriented vector
  sets can be perfectly superposed. It is also worthwhile noting that
  the upper limit in the sum in (\ref{eq:LeastSquaresQuat}) can be
  changed from 3 to 2, since two linearly independent vectors with the
  same origin, here $\m{t}_{j}$ and $\m{n}_{j}$, suffice to define a
  rigid body.

\item Extract $\m{n}_{j}$ and $\phi_{j}$ from the quaternion
  parameters $Q_{j}$. This can be easily achieved by expoiting the
  relations
  \begin{equation}
    \label{eq:qnPhi}
    \left.
      \begin{array}{lll}
        q_{0}&=&\cos(\phi/2) \\
        q_{1}&=&\sin(\phi/2)n_{x} \\
        q_{2}&=&\sin(\phi/2)n_{y}\\
        q_{3}&=&\sin(\phi/2)n_{z}
      \end{array}
    \right\}
  \end{equation}
  Here and in the following the index~$j$ is dropped. Several cases
  have to be considered.  If
  $\sqrt{q_{1}^{2}+q_{2}^{2}+q_{3}^{2}}>\epsilon$, where $\epsilon$
  depends on the machine precision of the computer being used, we
  compute a ``tentative rotation axis''
  \begin{equation}
    \label{eq:nTest}
    \m{n}_{t} = \frac{1}{\sqrt{q_{1}^{2}+q_{2}^{2}+q_{3}^{2}}}
    \left(\begin{array}{c}q_{1} \\q_{2} \\q_{3}\end{array}\right).
  \end{equation}
  Then we check if $\m{t}\cdot\m{n}_{t}\ge 0$. If this is the case we
  set
  \begin{align}
    \label{eq:n1}
    \m{n}&=\m{n}_{t},   \\
    \label{eq:phi1}
    \phi &=2\arccos(q_{0}).
  \end{align}
  In case that $\m{t}\cdot\m{n}_{t}< 0$ we set
  \begin{align}
    \label{eq:n2}
    \m{n}&=-\m{n}_{t},   \\
    \label{eq:phi2}
    \phi &=2\arccos(-q_{0}).
  \end{align}
  This corresponds to replacing $Q\to -Q$ before evaluating $\m{n}$
  and $\phi$ according to (\ref{eq:n1}) and (\ref{eq:phi1}). Such a
  replacement is possible since the elements of $\m{D}(Q)$ are
  homogeneous functions of order two in the quaternion parameters,
  such that $\m{D}(Q)=\m{D}(-Q)$.

  For the sake of completeness, we \gerald{finally mention} the case
  that $\sqrt{q_{1}^{2}+q_{2}^{2}+q_{3}^{2}}\le\epsilon$, which
  corresponds to a pure translation and cannot occur in our
  application to protein backbones. In this case one would set
  $\phi=0$ and $\m{n}=\m{t}/|\m{t}|$.

\item Using the parameters $\{\m{n}_{j},\phi_{j}\}$ and defining the
  positions $\m{R}_{j}$ to be the rotation centers,
  $\m{x}^{(c)}=\m{R}_{j}$, compute for $j=1,\ldots,N-1$
  \begin{enumerate}
  \item the positions $\m{s}^{(c)}_{j}$ on the local screw axes
    according to relation (\ref{eq:Sperp}),
  \item the local helix radii according to relation (\ref{eq:rho}).
  \end{enumerate}
\end{enumerate}

\subsubsection{Regularity of PSSEs}
To quantify the regularity of PSSEs, we introduce the distance measure
\begin{equation}
  \label{eq:RegularityMeasure}
  \delta(j) = \left|\m{s}^{(c)}_{j}+\m{t}^{\|}_{j}-\m{s}^{(c)}_{j+1}\right|,
  \quad j=1,\ldots,N-2,
\end{equation}
where $\m{t}^{\|}_{j}=\m{n}\cdot\m{t}_{j}$. For an ideal PSSE, where
all consecutive Frenet frames are related by the same screw motion,
$\delta(j)$ is strictly zero. This measure of non-ideality deviates
from the ``straightness'' parameter in the ScrewFit
algorithm~\cite{Kneller:2006p13}, which is defined as
$\sigma_{j}=\gm{\mu}_{j+1}\cdot\gm{\mu}_{j}$ with
$\gm{\mu}_{j}=\m{s}^{(c)}_{j+1}-\m{s}^{(c)}_{j}$, and which defines
ideality of PSSEs through the cosine of the angle between subsequent
local screw axes.

\begin{figure}[t]
  \begin{center}
    \includegraphics[scale=0.7]{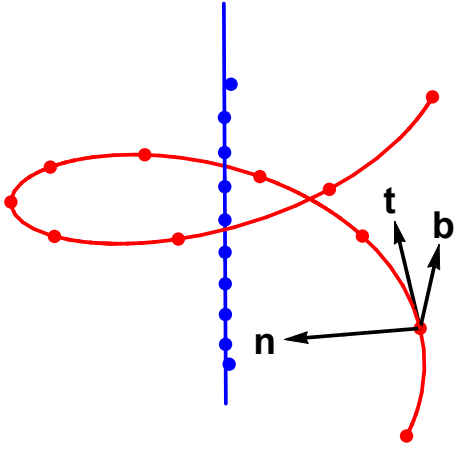}
    \caption{Frenet frame $\{\m{t},\m{n},\m{b}\}$ at one point of the
      helicoidal curve defined in Eq. (\ref{eq:HelicoidalCurve}) (red
      solid line).  Setting $R=1$ and $h=0.3$, the latter is shown for
      one turn, together with $N=11$ equidistantly spaced sampling
      points (red points). The blue line is the helix axis and the
      blue points correspond to the rotation centers $\m{s}^{(c)}_{j}$
      (\mbox{$j=1,\ldots N-1$}). The figure has been produced with the
      Mathematica software~\cite{Mathematica10}.}
    \label{fig:FrenetFrame}
  \end{center}
\end{figure}
\subsection{Numerical test}
\label{sec:NumericalTest}
To test the numerical construction of Frenet frames, we consider a
\gerald{perfect} helicoidal curve and compare the exact Frenet frames
with the corresponding numerical approximations. The parametric
representation of the curve is
\begin{equation}
  \label{eq:HelicoidalCurve}
  \m{r}(\lambda)=\rho\cos(\lambda) \,\m{e}^{(x)}+\rho\sin(\lambda) \,\m{e}^{(y)}
  +h\lambda\,\m{e}^{(z)},
\end{equation}
where $\rho>0$ is the radius of the helix and its pitch is $p=h/2\pi$.
Fig.~\ref{fig:FrenetFrame} shows the form of the
curve~(\ref{eq:HelicoidalCurve}) for one complete turn (red line),
setting $R=1$ and $h=0.3$ in arbitrary length units.  Defining the
matrix
$\m{F}(\lambda)=(\m{t}(\lambda),\m{n}(\lambda),\m{b}(\lambda))$, it
follows from (\ref{eq:HelicoidalCurve}) that
\begin{equation}
  \label{eq:FrenetExact}
  \m{F}(\lambda)=\left(
    \begin{array}{ccc}
      -\frac{R \sin (\lambda )}{\sqrt{h^2+R^2}} & -\cos (\lambda
      ) & \frac{h \sin (\lambda )}{\sqrt{h^2+R^2}} \\\noalign{\medskip}
      \frac{R \cos (\lambda )}{\sqrt{h^2+R^2}} & -\sin (\lambda )
      & -\frac{h \cos (\lambda )}{\sqrt{h^2+R^2}} \\\noalign{\medskip}
      \frac{h}{\sqrt{h^2+R^2}} & 0 & \frac{R}{\sqrt{h^2+R^2}} \\
    \end{array}
  \right).
\end{equation}
Using the method described in Section~\ref{sec:FrenetFrames}, we
construct numerical approximations $\tilde{\m{F}}(\lambda_{j})$ of the
Frenet bases (\ref{eq:FrenetExact}) at $N=11$ equidistant sampling
points, $\m{R}_{j}$, which are shown as red dots in
Fig.~\ref{fig:FrenetFrame}.  From these Frenet bases we construct the
axis points $\m{s}^{(c)}_{j}$ (blue dots), which are shown together
with the exact screw axis (blue line). For the first and the last axis
point one notices a visible offset from the latter.  \gerald{We
  quantify the error of the numerically computed Frenet bases,
  $\tilde{\m{F}}(\lambda_{j})$, as}
\begin{equation}
  \label{eq:OverlapError}
  \epsilon(j) = \sqrt{\text{tr}\left\{\gm{\Delta}(j)^{T}\cdot\gm{\Delta}(j)\right\}},
\end{equation}
where
\begin{equation}
  \gm{\Delta}(j)=\tilde{\m{F}}(\lambda_{j})^{T}\cdot\m{F}(\lambda_{j})-\m{1}.
\end{equation}
For a perfect overlap of $\tilde{\m{F}}(\lambda_{j})$ and
$\m{F}(\lambda_{j})$ one should have
$\tilde{\m{F}}(\lambda_{j})^{T}\cdot\m{F}(\lambda_{j})=\m{1}$,
\gerald{such that $\epsilon(j)=0$.} We note that $\epsilon(j)$ is the
Frobenius norm~\cite{Golub:1996} of~$\gm{\Delta}(j)$.
Fig.~\ref{fig:FrenetFrameError} shows $\epsilon(j)$ corresponding to
the Frenet basis in Fig.~\ref{fig:FrenetFrame} and confirms the slight
offset of the first and last axis point from the ideal screw axis.
\begin{figure}
  \begin{center}
    \includegraphics[scale=0.6]{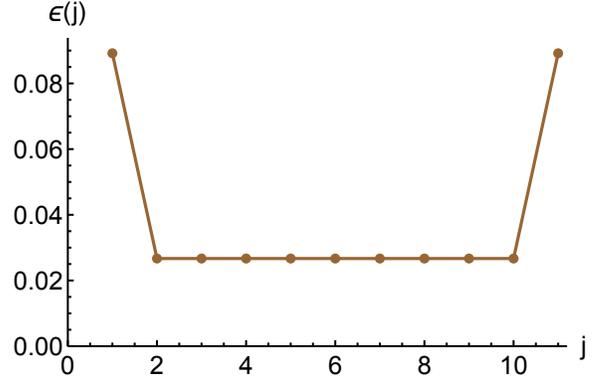}
    \caption{Overlap error~(\ref{eq:OverlapError}) for the bases
      $\tilde{\m{F}}(\lambda_{j})$ and $\m{F}(\lambda_{j})$ at the red
      points in Fig.~\ref{fig:FrenetFrame}.}
    \label{fig:FrenetFrameError}
  \end{center}
\end{figure}

\section{Applications}
\label{sec:Applications}

In the following we consider two applications of the coarse-grained
model for protein secondary structure, which has been described in the
previous section and which will be referred to as ScrewFrame in the
following.  The first application concerns the construction of a tube
model for a protein from the ScrewFrame parameters and in the second
application, these parameters are used for a comparative study of
ScrewFrame and DSSP for secondary structure assignment.
\begin{figure}
  \begin{center}
    \includegraphics[scale=0.7]{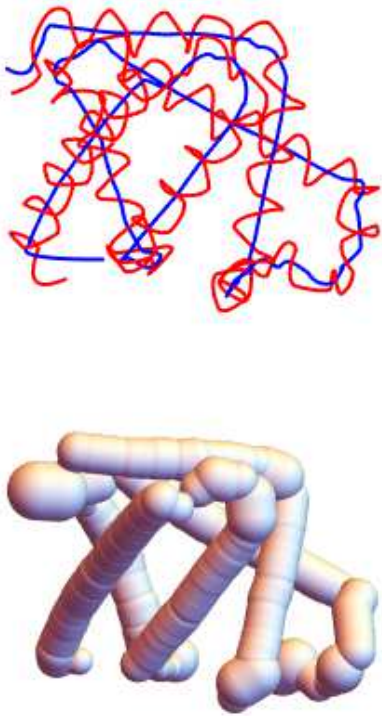}
    \caption {\textbf{Top:} $C_{\alpha}$-curve (red) of
      myoglobin (PDB code 1A6G) and B-spline curve (blue) linking the
      screw motion centers $\{\m{s}^{(c)}_{j}\}$.  \textbf{Bottom:}
      Tube representation of the $C_{\alpha}$-curve. The local tube
      radii equal the respective helix radii $\{\rho_{j}\}$ of the
      screw motions linking the Frenet frames $j$ and $j+1$
      ($j=1,\ldots, N-1$).  The figure has been produced with the
      Mathematica software~\cite{Mathematica10}.}
    \label{fig:ScrewAxis}
  \end{center}
\end{figure}
\subsection{Tube representation of a protein}
As a first application we consider the ScrewFrame model for myoglobin,
which is an oxygen-binding protein in muscular tissues. Myoglobin is
composed of 151 amino acids which fold into a globular form and the
dominant PSSEs are $\alpha$-helices. For our demonstration we use the
crystallographic structure 1A6G of the Protein Data
Bank~\cite{Kirchmair:2008p252}. The red and blue line in the upper
part of Fig.~\ref{fig:ScrewAxis} display, respectively, the space
curve defined by the positions $\m{R}_{j}$ of the $C_{\alpha}$-atoms
and the space curve linking the corresponding screw motion centers
$\m{s}^{(c)}_{j}$.  Both space curves are constructed by a piecewise
polynomial interpolation of second order~\cite{Mathematica10}. The
blue line indicates the global fold of the protein, where ideal PSSEs
appear simply as straight segments. In the following we refer to this
line as the protein screw axis.  It plays the same role as the
``overall protein axis'' in the P-Curve
algorithm~\cite{Sklenar:1989vv}, although its construction is
different. The lower part of the figure shows the corresponding ``tube
model'', where the axis of the tube equals the protein screw axis and
the local tube radius corresponds to the radius of the local screw
motion.  As in the original ScrewFit algorithm, the screw radius
allows for a discrimination of different types of PSSEs (see
Table~\ref{Tab:RhoModel}).
\begin{table}[tdp]
  \begin{center}\footnotesize
    \begin{tabular}{|l|l|l|l|l|l|}
      \hline
      &$\alpha$-helix
      &$\beta$-strand $\uparrow \uparrow$
      &$\beta$-strand $\uparrow \downarrow$
      &3-10 helix 
      &$\pi$-helix\\
      \hline
      \footnotesize{ScrewFit}   
      & 0.165          & 0.061                    & 0.051                        & 0.122      &0.165\\
      \footnotesize{ScrewFrame}  
      & 0.227          & 0.098                    & 0.080                        & 0.187      &0.227\\
      \hline
    \end{tabular} 
    \caption{\label{Tab:RhoModel} Screw radii in nm for standard model
      structures generated \gerald{with}
      Chimera~\cite{Pettersen:2004kh}. Since ScrewFit uses the
      $C$-atoms in the peptide planes as reference points for the
      (pure) rotations, whereas as ScrewFrame uses the
      $C_{\alpha}$-atoms, the radii determined by ScrewFit are
      systematically smaller than those obtained from ScrewFrame.}
  \end{center}
\end{table}
\begin{figure}[b]
  \begin{center}
    \includegraphics[scale=0.65]{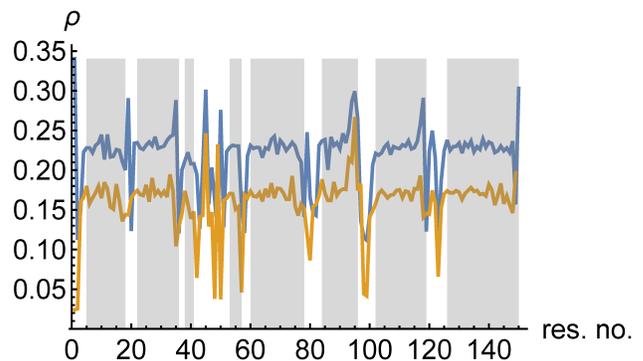}
    \caption {The radius $\rho$ for the ScrewFrame representation
      (blue line) of myoglobin (PDB code 1A6G) as a function of the
      residue number and the corresponding values for ScrewFit (brown
      line).  The light gray stripes indicate the $\alpha$-helices
      found by DSSP.}
    \label{fig:ScrewRho}
  \end{center}
\end{figure}
Fig.~\ref{fig:ScrewRho} displays this quantity for myoglobin as a
function of the residue number (blue line) and, for comparison, the
corresponding values for the ScrewFit algorithm (brown line). The
light gray stripes indicate $\alpha$-helices found by the DSSP
algorithm. The comparison of the results with the ScrewFit analysis of
the same protein structure shows that both methods indicate
$\alpha$-helices in the same place, in close agreement with DSSP. Here
it must be observed that the definition of the screw radii is not the
same for ScrewFit and ScrewFrame. The rotation centers in the ScrewFit
algorithm are the $C$-atoms in the $C-O-N$-peptide planes, whereas the
$C_{\alpha}$-atoms are used for ScrewFrame. For an ideal
$\alpha$-helix the corresponding radii are 0.165~nm and 0.227~nm,
respectively (see Table~\ref{Tab:RhoModel}). Fig.~\ref{fig:Regularity}
shows the regularity measure~(\ref{eq:RegularityMeasure}) which plays
an important role in the attribution of secondary structure elements
to be discussed in the following section.
\begin{figure}[t]
  \begin{center}
    \includegraphics[scale=0.65]{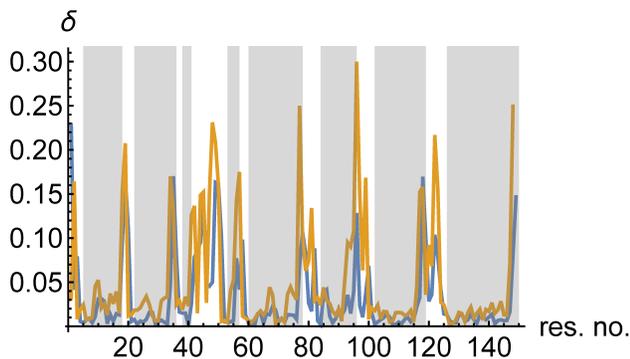}
    \caption {The regularity measure (\ref{eq:RegularityMeasure}) for
      the ScrewFrame representation of myoglobin (PDB code 1A6G) as a
      function of the residue number and the corresponding values for
      ScrewFit (brown line). The light gray stripes indicate the
      $\alpha$-helices found by DSSP.}
    \label{fig:Regularity}
  \end{center}
\end{figure}

\subsection{Analysis of the ASTRAL database}
In order to compare our C$_{\alpha}$ based helicoidal analysis with
the original ScrewFit method based on peptide
planes~\cite{Kneller:2006p13,Calligari:2012eja}, we applied both
methods to the ``all $\alpha$ and ``all $\beta$'' categories of the
ASTRAL subset of the SCOPe database~\cite{Fox:2013in}, using the
ASTRAL SCOPe 2.04 subset with less than 40\% sequence identity. In
order to be able to work efficiently with such a large collection of
protein structures, we constructed an ActivePaper~\cite{ActivePapers}
containing the structures of the ASTRAL entries in MOSAIC
format~\cite{Hinsen:2014cb}.  This file is available for
download~\cite{ASTRAL-AP}. In addition to the ASTRAL database of real
protein structures, we use ideal secondary-structure elements
($\alpha$-helix, $\pi$-helix, $3-10$-helix, parallel and anti-parallel
$\beta$-strands) for polyalanine, which were constructed using the
program Chimera~\cite{Pettersen:2004kh}.

\begin{figure}[t]
  \begin{center}
    \includegraphics[scale=0.43]{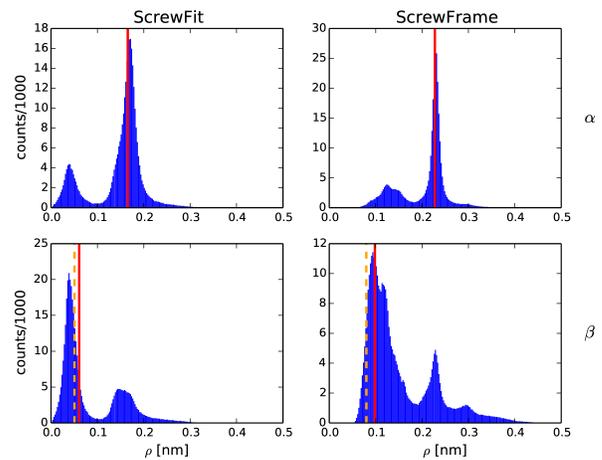}
    \caption{The helix radius $\rho$ for the all-$\alpha$ (top) and
      the all-$\beta$ structures (bottom), using the ScrewFit (left)
      and ScrewFrame (right) methods. Note that the ScrewFit radius is
      based on the C-atoms, whereas the ScrewFrame radius corresponds
      to the C$_{\alpha}$-atoms, which explains the different values.
      The vertical lines indicate the values for ideal
      secondary-structure elements. For $\beta$-strands, there are two
      ideal values, one for parallel (red, drawn-out) and one for
      antiparallel (orange, dashed) strands.}
    \label{fig:rho}
  \end{center}
\end{figure}
\begin{figure}[h!]
  \begin{center}
    \includegraphics[scale=0.43]{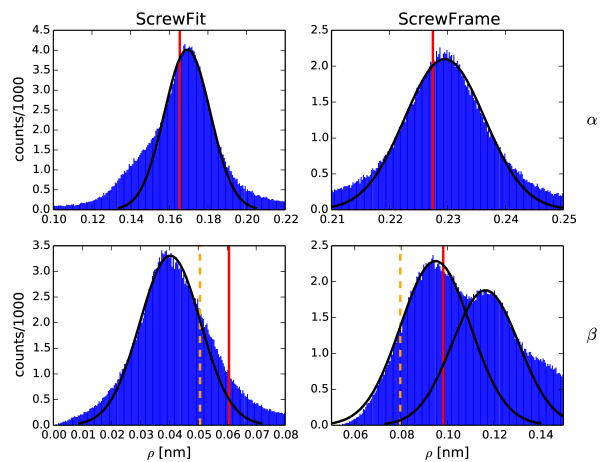}
    \caption{The helix radius $\rho$ around the ideal-$\alpha$ value
      for the all-$\alpha$ subset (top) and around the ideal-$\beta$
      values for the the all-$\beta$ structures (bottom), using the
      ScrewFit (left) and ScrewFrame (right) methods. The vertical
      lines indicate values for ideal secondary-structure elements, as
      in Fig.~\ref{fig:rho}. \gerald{The} Gaussian distributions
      fitted to the peaks are \gerald{drawn} in black, their
      parameters are given in Table~\ref{Tab:RhoDistribution}. The
      $\beta$ distribution for ScrewFrame can be well described as a
      superposition of two Gaussian distributions, corresponding to
      parallel and antiparallel strands. \gerald{The ScrewFit method
        cannot resolve this difference.}}
    \label{fig:rho-detail}
  \end{center}
\end{figure}

We also compare to DSSP secondary structure assignments for this
database, using our own implementation of the DSSP algorithm which
follows the description in the original publication
\cite{Kabsch:1983wk} but, like the current version~2 of the DSSP
software~\cite{DSSP-Software}, computes an ideal position for the
backbone hydrogen positions instead of using experimental values, even
if the latter are available.

As a first step, we compute ScrewFit and ScrewFrame parameters for all
structures in the all-$\alpha$ and all-$\beta$ subsets of the ASTRAL
database. In order to avoid inaccuracies introduced by the third-order
\gerald{approximations given by
  Eqs.~(\ref{eq:third-order-1})--(\ref{eq:third-order-4})}, we do not
compute Frenet frames for the first and last residue of each
chain. For structures with missing residues, we compute the parameters
for each continuous chain segment separately. Since the input
structures are dominated by $\alpha$-helices and $\beta$-strands,
respectively, we expect the distribution of our parameters to show
clear peaks that correspond to these secondary structure elements.

The most important helix parameter for secondary structure description
is the helix radius $\rho$, whose distribution in the ASTRAL database
is shown in Fig.~\ref{fig:rho}. The vertical lines show for comparison
the values for ideal $\alpha$-helices and $\beta$-strands. For the
$\beta$-strands, the red drawn-out lines stand for parallel and the
orange dashed lines for antiparallel strands.  A more detailed view is
given in Fig.~\ref{fig:rho-detail}, which shows only the region around
the dominant peak for each histogram, together with Gaussian
distributions fitted to the peaks. The peaks are rather well described
by a Gaussian, and the ScrewFrame method even allows to resolve the
difference between parallel and antiparallel $\beta$-strands.

Whereas the average $\rho$ value for $\alpha$-helices is close to the
value for an ideal helix, this is not the case at all for
$\beta$-strands.  This can be understood by looking at the
distribution of the number of amino acids per full turn, $\tau$, shown
in Fig.~\ref{fig:tau}. Since the rotation angle is by definition in
the interval $[-\pi \ldots \pi]$, the minimal value of $\tau$
is~2. This is also the value that describes an ideal $\beta$-strand,
which is a flat structure. Any deviation from the ideal $\beta$-strand
has a larger $\tau$, and because $\rho$ and $\tau$ are not independent
(the length of the curve arc linking two neighboring C$_\alpha$ atoms
is nearly constant), the deviation in $\rho$ from the ideal value is
asymmetric as well.
\begin{figure}[t]
  \begin{center}
    \includegraphics[scale=0.43]{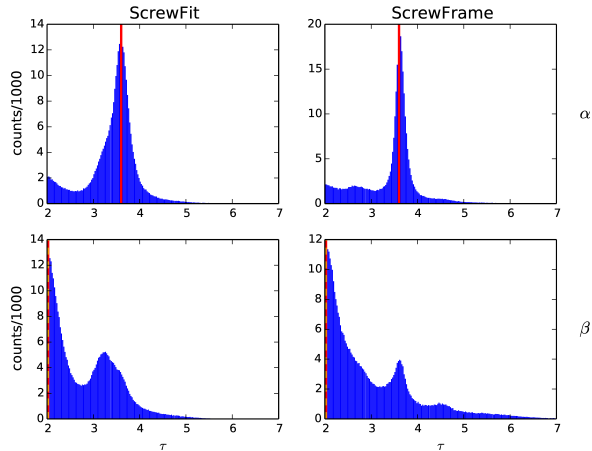}
    \caption{The number of amino acid residues per full turn, $\tau$,
      for the all-$\alpha$ \gerald{(top)} and the all-$\beta$
      structures \gerald{(bottom)} using the ScrewFit \gerald{(left)}
      and ScrewFrame \gerald{(right)} methods.  The theoretical
      minimal value of $\tau=2$ is very close to the observed value
      for $\beta$-sheets.}
    \label{fig:tau}
  \end{center}
\end{figure}
\begin{figure}[h]
  \begin{center}
    \includegraphics[scale=0.43]{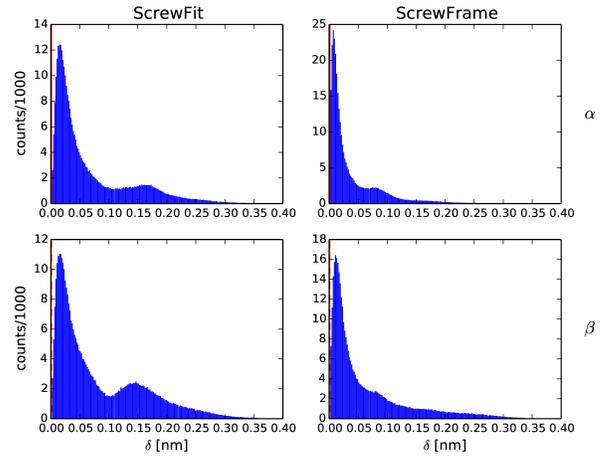}
    \caption{The regularity measure $\delta$ defined in
      Eq.~(\ref{eq:RegularityMeasure}) \gerald{for the all-$\alpha$
        and the all-$\beta$ subset of the ASTRAL data base (top and
        bottom, respectively)}.}
    \label{fig:delta-r}
  \end{center}
\end{figure}

The regularity measure $\delta$, defined in
Eq.~(\ref{eq:RegularityMeasure}), is shown in Fig.~\ref{fig:delta-r}.
It shows that the ScrewFrame secondary structure elements are more
regular than those identified by ScrewFit, in particular for
structures dominated by $\alpha$-helices.  We do not show here the
distributions of the other parameters defined in the initial ScrewFit
publication~\cite{Kneller:2006p13}, but they are included in the
electronic supplementary material. \gerald{We note that the parameter
  distributions are in general} narrower and thus better defined for
ScrewFrame than for ScrewFit. We attribute this fact to fluctuations
in the orientations of the peptide plans that have no impact on the
C$_{\alpha}$ geometry.
\begin{figure}
  \begin{center}
    \includegraphics[scale=0.35]{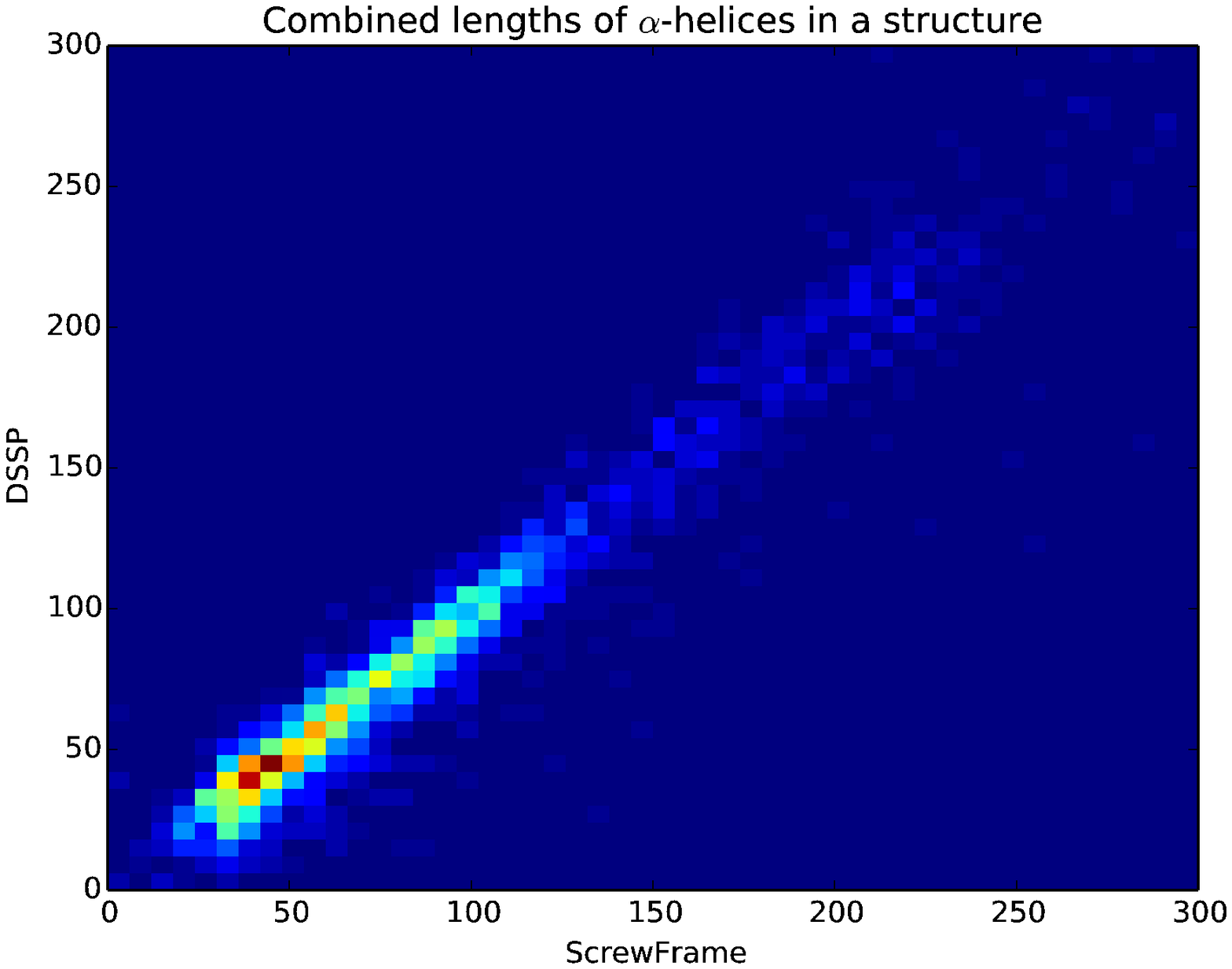}\\
    \includegraphics[scale=0.35]{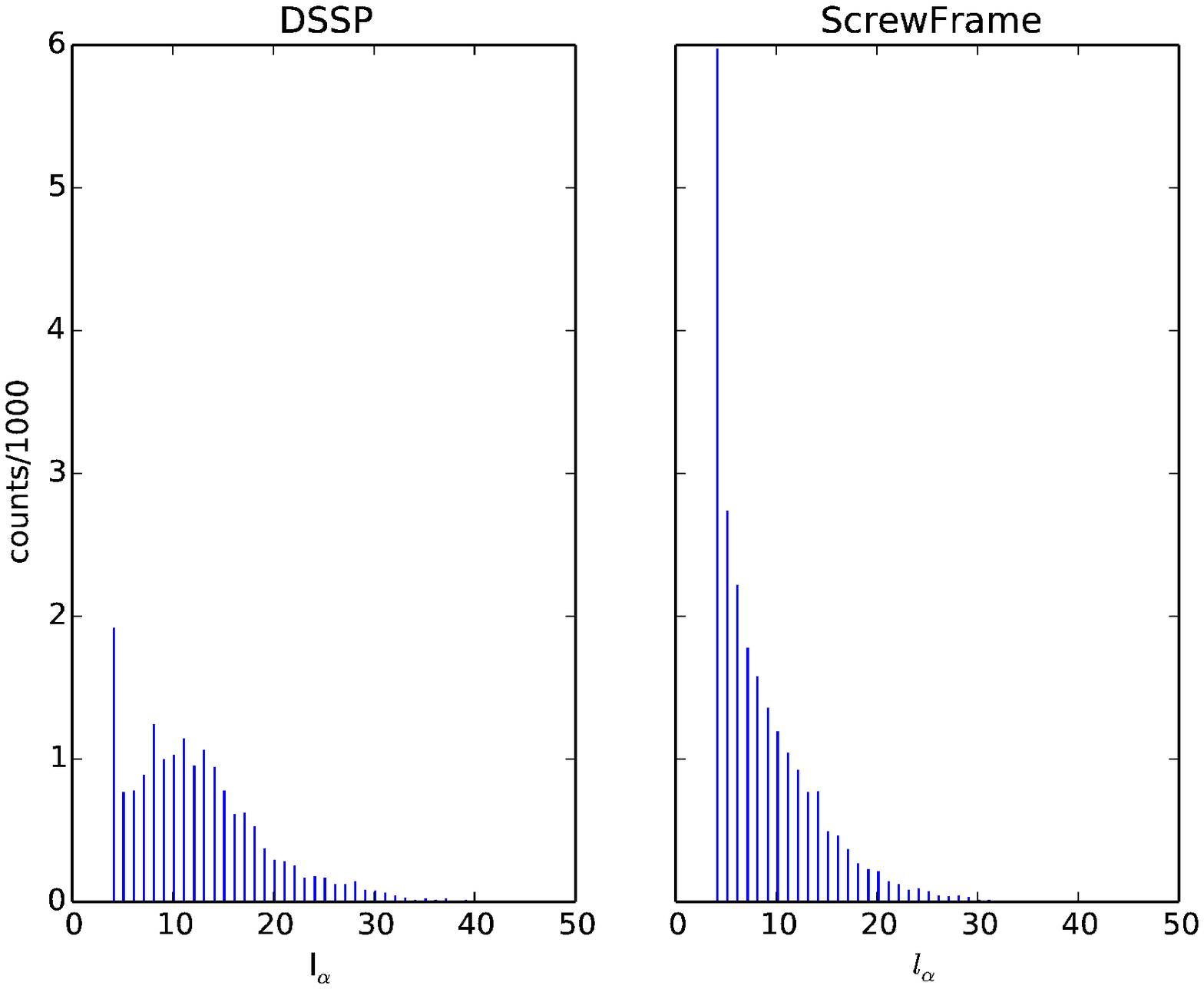}
    \caption{\textbf{Top:} A two-dimensional histogram comparing the
      total number of residues inside $\alpha$-helices as identified
      by ScrewFrame and DSSP. The strong localization of the
      distribution around the diagonal shows the similarity between
      these two assignments. \textbf{Bottom:} The distribution of the
      lengths of identified $\alpha$-helices, left for DSSP, right for
      ScrewFrame. The fatter tail for DSSP and the larger number of
      short helices for ScrewFrame are due to the fact that ScrewFrame
      breaks up strongly deformed helices into several pieces, whereas
      DSSP considers them a single helix.}
    \label{fig:alpha-comparison}
  \end{center}
\end{figure}

We use the Gaussian distributions shown in Fig.~\ref{fig:rho-detail}
as the basis for defining secondary-structure elements. We define an
$\alpha$-helix as a sequence of at least four consecutive C$_\alpha$
atoms whose screw transformations satisfy
\begin{eqnarray}
  \frac{|\rho-\mu_\rho|}{\sigma_\rho} &<& 3 \\
  \delta &<& 0.02\, \mbox{nm}
\end{eqnarray}
where $\mu_\rho$ and $\sigma_\rho$ are the mean value and standard
deviation of the Gaussian distribution for the $\alpha$ peak in
Fig.~\ref{fig:rho-detail}. The numerical values of these parameters
are shown in Table~\ref{Tab:RhoDistribution}.
\begin{table}[h!]
  \begin{center}\footnotesize
    \begin{tabular}{|l|c|c|c|}
      \hline
      & $\alpha$-helix & $\beta$-strand $\uparrow \uparrow$ 
      & $\beta$-strand $\uparrow \downarrow$ \\
      \hline
      $\mu_\rho$     & 0.230          & 0.116                    & 0.095   \\
      $\sigma_\rho$  & 0.007          & 0.014                    & 0.015   \\
      \hline
    \end{tabular} 
    \caption{\label{Tab:RhoDistribution} The parameters of the
      Gaussians fitted to the peaks in the distributions of the
      ScrewFrame parameter $\rho$ (see Fig.~\ref{fig:rho-detail}). All
      values are in units of nm.}
  \end{center}
\end{table}    

We define a $\beta$-strand as a segment of consecutive C$_\alpha$
atoms whose screw transformations satisfy
\begin{eqnarray}
  \min\left(\frac{|\rho-\mu_\rho^{(1)}|}{\sigma_\rho^{(1)}},
    \frac{|\rho-\mu_\rho^{(2)}|}{\sigma_\rho^{(2)}}\right) &<& 1 \\
  \delta &<& 0.08\, \mbox{nm}
\end{eqnarray}
where $\mu_\rho^{(1/2)}$ and $\sigma_\rho^{(1/2)}$ are the mean values
and standard deviations of the Gaussian distributions for the parallel
and antiparallel $\beta$ peaks in Fig.~\ref{fig:rho-detail}. The
numerical parameters in these definitions were chosen to make our
definitions match the secondary structure assignments made by the DSSP
method.

There is a fundamental difference between our approach and the DSSP
method for defining $\beta$-strands. The ScrewFrame approach looks for
a regular structure along the peptide chain, whereas the DSSP method
identifies hydrogen bonds between the strands that make up a
$\beta$-sheet.  ScrewFrame thus finds individual strands, which can be
paired up to identify sheets in a separate step. A strand must consist
of at least three consecutive residues in order to be considered
regular; in fact, the regularity measure $\delta$ is defined in terms
of the difference of two consecutive screw transformations, each of
which connects two residues. DSSP needs to look at two strands
simultaneously in order to identify $\beta$ structures, but has no
minimal length condition and in fact admits $\beta$-sheets as mall as
a single h-bonded residue pair. \gerald{For practically relevant
  $\beta$-sheets in real protein structures, these differences are,
  however,} not important, but they must be understood for
interpreting the following comparison between the two methods.

\begin{figure}[t]
  \begin{center}
    \includegraphics[scale=0.35]{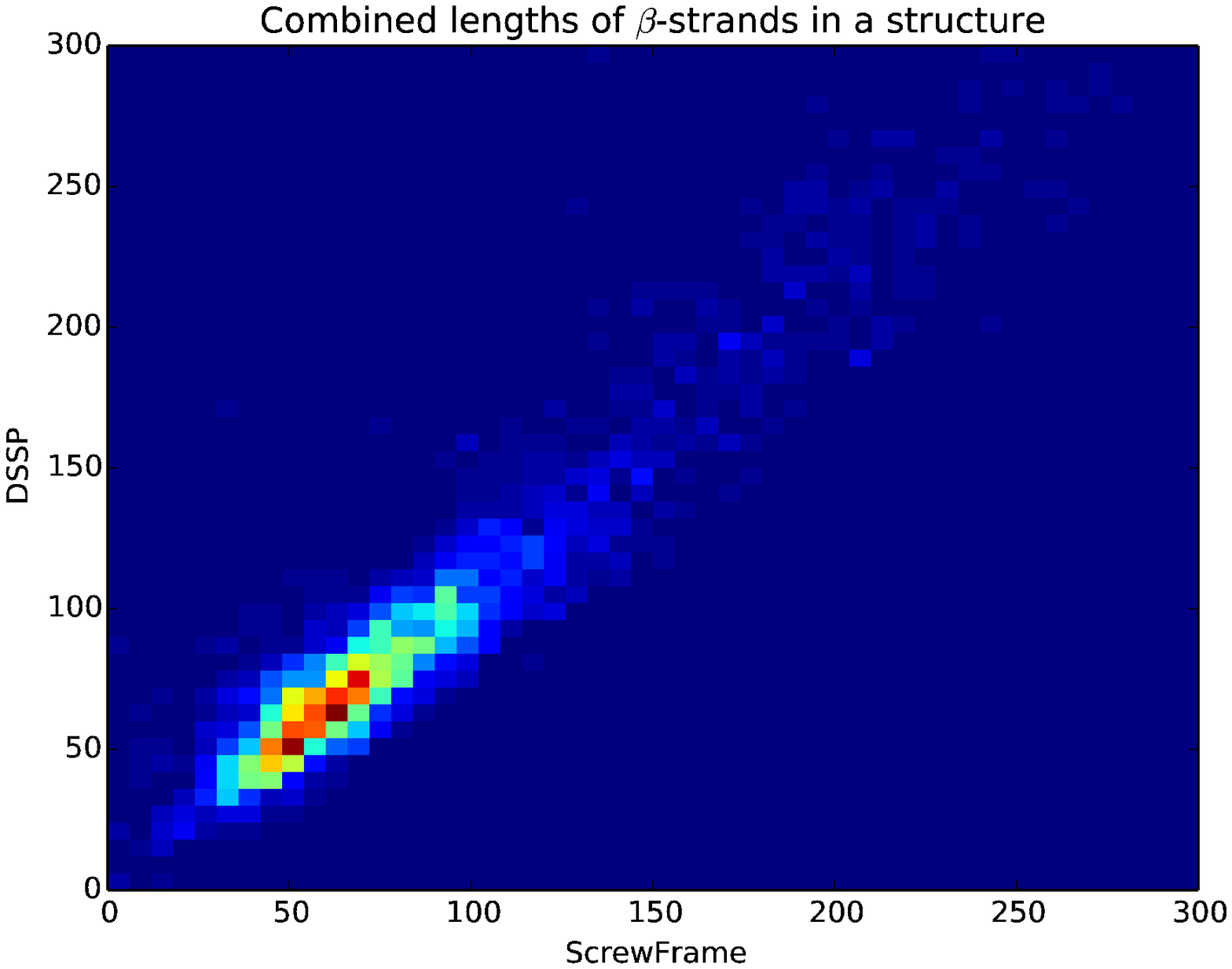}\\
    \includegraphics[scale=0.35]{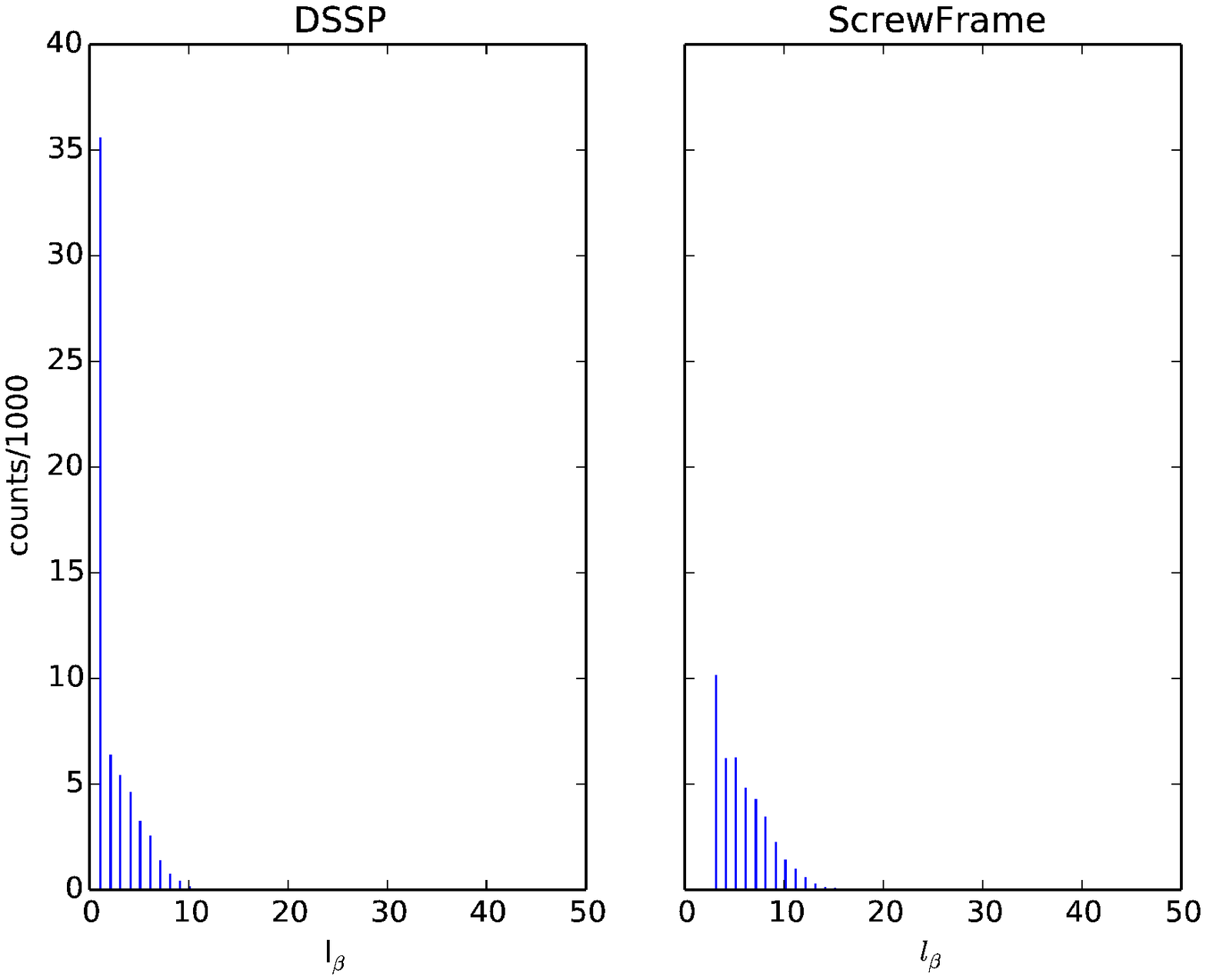}
    \caption{\textbf{Top: }A two-dimensional histogram comparing the
      total number of residues inside $\beta$-strands as identified by
      ScrewFrame and DSSP. The strong localization of the distribution
      around the diagonal shows the similarity between these two
      assignments. \textbf{Bottom:} The distribution of the lengths of
      identified $\beta$-helices, left for DSSP, right for
      ScrewFrame. The peak at very short strands in the DSSP
      distribution is absent from the ScrewFrame results because
      ScrewFrame needs at least three consecutive residues to
      recognize a regular structure.}
    \label{fig:beta-comparison}
  \end{center}
\end{figure}

A one-to-one comparison of secondary structure elements from two
different assignment methods is not of particular interest, because an
exact match is the exception rather than the rule. The inherent
fuziness of secondary structure definitions leads to arbitrary choices
and thus inevitable differences. The most frequent deviation between
two assignments is the end points of secondary structure elements,
where a difference of one or two residues is common and acceptable.
Another frequent deviation concerns deformed secondary structure
elements, which one method may identify as a single element whereas
another one recognizes it as multiple distinct elements.

We therefore chose a statistical comparison to compare the ScrewFrame
results to those of DSSP, which is shown in
Figs.~\ref{fig:alpha-comparison} for $\alpha$-helices
and~\ref{fig:beta-comparison} for $\beta$-strands.  We consider two
quantities: (1) the total number of residues of a given structure
which are inside a recognized secondary-structure element, and (2) the
length of each individual secondary-structure element. We compute the
first quantity for both methods and show their joint distribution
(upper plot in the two figures). For the vast majority of structures,
the two residue counts are close to equal, which means that neither
method yields systematically more or longer secondary-structure
elements than the other. The lower plots show the distributions of the
lengths of individual secondary-structure elements. For
$\alpha$-helices, DSSP has a fatter tail (helices of length~20 or
more), whereas ScrewFrame identifies a larger number of short helices.
The reason for these differences is that ScrewFrame tends to split up
kinked helices which DSSP identifies as single units. For
$\beta$-strands, we notice that DSSP identifies many more very short
elements. This is due to the different definitions: a single
$\beta$-type hydrogen bond is sufficient to define a $\beta$-sheet in
DSSP, but ScrewFrame requires at least three consecutive residues to
identify any regular structure.

\section{Conclusion and Outlook}
\label{sec:Conclusions}

We have presented a generalization of the ScrewFit method for protein
structure assignment and description, which uses only the positions of
the $C_{\alpha}$-atoms along the protein backbone. As in the ScrewFit
approach, the global protein fold is described as a succession of
screw motions relating consecutive recurrent motifs along the protein
backbone, but the ``motifs'' are here the tripods (planes) formed by
the three (two) orthonormal vectors of the local Frenet bases to the
$C_{\alpha}$ space curve. Despite the fact that ScrewFrame uses less
information than ScrewFit, all standard PSSEs are recognized on the
basis of thresholds for the local screw radii and a suitably defined
regularity measure. ScrewFrame even permits to distinguish between
parallel and antiparallel $\beta$-strands, which the classical
ScrewFit method fails to do. A thorough comparison with the commonly
used DSSP method on the assignment of PSSEs in the ASTRAL database
shows that both methods yield very similar results for the total
amount of PSSEs. ScrewFrame tends, however, to break long helices into
smaller pieces, such that the length distribution of PSSEs is
different. Due to the minimalistic character of the geometrical model
for protein folds, the evaluation of the ScrewFrame model parameters
is very efficient. This allows for working with protein structure
databases and for analyzing simulated molecular dynamics trajectories
of proteins.  ScrewFrame may also be used a starting point for the
development of minimalistic models for protein structure and dynamics,
similar to the wormlike chain model~\cite{Doi:1986}, which has been
successfully applied to DNA~\cite{Marko:1995wn}. As already mentioned,
our method can also be used to analyze dynamical processes, such as
the folding and unfolding of peptides~\cite{Spampinato:2014wd}
\gerald{and it can describe the fold of intrinsically disordered
  proteins.}
 
\vspace{5mm} An ActivePaper~\cite{ActivePapers} containing all the
software, input datasets, and results from this study is available as
supplementary material. The datasets can be inspected with any
HDF5-compatible software, e.g. the free HDFView.\cite{HDFView} Running
the programs on different input data requires the ActivePaper
software~\cite{ActivePapers}.

\vfill


\end{sloppy}
\end{document}